%% file: iclr2025.tex
\documentclass{article} 
\usepackage{iclr2025,times}
\usepackage{graphicx} 
\usepackage{adjustbox}
\usepackage{hyperref}
\usepackage{booktabs,multirow}
\usepackage{enumitem}
\usepackage{comment}
\usepackage{amsmath}
\usepackage{amssymb}
\usepackage{physics}
\usepackage{tabularx}
\usepackage{float}

\input{math_commands.tex}

\usepackage{hyperref}
\usepackage{url}

\title{Modeling speech emotion with label variance and analyzing performance across speakers and unseen acoustic conditions}

\author{Vikramjit Mitra, Amrit Romana, Dung T. Tran \& Erdrin Azemi \\
Apple\\
Cupertino, CA 95014, USA \\
\texttt{\{vmitra,aromana,dung\_tran,eazemi\}@apple.com} \\
}

\iclrfinalcopy 
\begin{document}

\maketitle

\begin{abstract}
Spontaneous speech emotion data usually contain perceptual grades where graders assign emotion score after listening to the speech files. Such perceptual grades introduce uncertainty in labels due to grader opinion variation. Grader variation is addressed by using consensus grades as groundtruth, where the emotion with the highest vote is selected. Consensus grades fail to consider ambiguous instances where a speech sample may contain multiple emotions, as captured through grader opinion uncertainty. We demonstrate that using the probability density function of the emotion grades as targets instead of the commonly used consensus grades, provide better performance on benchmark evaluation sets compared to results reported in the literature. We show that a saliency driven foundation model (FM) representation selection helps to train a state-of-the-art speech emotion model for both dimensional and categorical emotion recognition. Comparing representations obtained from different FMs, we observed that focusing on overall test-set performance can be deceiving, as it fails to reveal the models generalization capacity across speakers and gender. We demonstrate that performance evaluation across multiple test-sets and performance analysis across gender and speakers are useful in assessing usefulness of emotion models. Finally, we demonstrate that label uncertainty and data-skew pose a challenge to model evaluation, where instead of using the best hypothesis, it is useful to consider the 2- or 3-best hypotheses.
\end{abstract}
\vspace{-4mm}
\section{Introduction}
\label{sec:intro}
\vspace{-2mm}
Speech-based emotion models aim to estimate the emotional state of a speaker from their speech utterances. Real-time speech-emotion models can help to improve human-computer interaction \cite{mitra2019leveraging, kowtha2020detecting} and facilitate health applications \cite{stasak2016investigation, niu2023capturing, provost2024emotion}. Speech emotion research has pursued two distinct definitions of emotion: (1) \textit{categorical emotions}: for example, fear, anger, joy, sadness, disgust, and surprise \cite{ekman1992argument}, and (2) \textit{dimensional emotions}: that represent emotion using a 3-dimensional model of \textit{Valence}, \textit{Activation} and \textit{Dominance} \cite{posner2005circumplex}. Early studies on speech emotion detection focused on acted or elicited emotions \cite{busso2008iemocap}, however, models trained with acted emotions often fail to generalize for spontaneous emotions \cite{douglas2005multimodal}. Recently, attention has been given to datasets with spontaneous emotions \cite{mariooryad2014building} where graders listen to each audio file and assign emotion labels. Such perceptual grading is difficult due to utterances containing mixed, shifting, subtle, or ambiguous emotions. To account for this, Mariooryad et al. have multiple graders review and grade each audio file. Traditionally, researchers addressed label variance by taking the grader consensus \cite{chou2024minority}. However, modeling such variance \cite{prabhu2022label, chou2024minority, tavernor2024whole} can be useful to account for audio samples that were perceptually difficult to annotate. In this work, we investigate training models with distributions of grader decisions for categorical emotions, instead of consensus grades, as the target. We hypothesize that modeling label uncertainty can help to improve the model's robustness because consensus grades fail to account for mixed, shifting, subtle, or ambiguous emotions. 

Recent studies have shown that pre-trained foundation model (FM) representations are useful for emotion recognition from speech \cite{srinivasan2021representation, mitra2022Speechemotion, mitra2023pre}. Given that the FMs may not have been trained with emotion labels, the final layer representations may not be optimal for emotion recognition. Earlier studies have investigated intermediate FM representations for various speech tasks \cite{alain2016understanding, mitra2020investigation, mitra2024pre, yang2024large}. In this work, we investigate saliency based FM layer selection for the downstream emotion modeling task. To summarize, in this work, we:
\vspace{-2mm}
\begin{enumerate}
\item Account label uncertainty through the use of categorical emotion pdf as targets.
\vspace{-1mm}
\item Explore saliency-driven intermediate FM layer representations for emotion recognition.
\vspace{-1mm}
\item Evaluate performance across speakers, gender and unseen acoustic conditions.
\vspace{-2mm}
\end{enumerate}

We observed that models that provide state-of-the-art (SOTA) results, may not generalize well across speakers and varying acoustic conditions. We found that having a diverse evaluation set along with a diverse evaluation metric is useful for model selection. We found that the traditional 1-best hypothesis used in emotion literature may get biased by the training data-skew, in which case 2- or 3-best hypotheses may be useful to account for speech samples containing multiple emotions.
\vspace{-2mm}
\section{Data}
\label{sec:data}
\vspace{-2mm}
We have used the MSP-Podcast dataset (ver. 1.11) \cite{mariooryad2014building, lotfian2017building} that contains $\approx 238$ hours of speech data spoken by English speakers ($N>1800$), consisting of $\approx 152K $ speaking turns. The speech segments contain single-speaker utterances with a duration of 3 to 11 seconds. The data contain manually assigned valence, activation and dominance scores and categorical emotions (9 categories) from multiple graders. Grader decisions for categorical emotions were converted to a pdf (reflecting the probability of each of the 9 emotions), which was used as the target for our model training. The data split is shown in Table \ref{tab:table1} in Appendix \ref{sec:data-split}. To make our results comparable to \cite{ghriss2022sentiment, srinivasan2021representation}, we report results on Eval1.6 and Eval1.11 (see Table \ref{tab:table1}). For evaluating model robustness, we have added noise to the MSP test-set at SNR levels 15 dB and 5dB (see $Eval_{15dB}$ and $Eval_{5dB}$ in Table \ref{tab:table1}, Appendix \ref{sec:data-split}). We report categorical emotion recognition performance on six emotions: neutral, happy, angry, sad, contempt and surprise. We have used CMU-Mosei, \cite{zadeh2018mosei} and a 5 hour in-house conversational speech data from 85 speakers for cross-corpus speech emotion recognition analysis. 

\vspace{-2mm}
\section{Representations}
\label{sec:rep}
\vspace{-2mm}
We explore speech embeddings as features to a TC-GRU model (see Figure \ref{fig:fig1}). We use the following pre-trained models to generate those embeddings: 
(i) \textbf{HuBERT} large \cite{hsu2021hubert}, a transformer based acoustic model, pre-trained on 60K hours of Libri-light speech data, generating 1024-dimensional embedding. 
(ii) \textbf{WavLM} large \cite{chen2022wavlm}, a transformer based acoustic model, generating 1024 dimensional embedding. WavLM has been pre-trained on 60K hours of Libri-light, 19K hours of GigaSpeech and 25K hours of VoxPopuli.
(iii) \textbf{Whisper} medium \cite{radford2023robust} acoustic model that generates 1024 dimensional embeddings from 24 transformer encoder layers. Whisper is trained with 680K hours of noisy and diverse speech data from the web.

Motivated by \cite{mitra2024investigating, mitra2024pre} we explore obtaining layer-saliency to obtain the optimal FM layer representation for emotion modeling. 
Let the $N$ dimensional representation from the $k^{th}$ layer of a FM for an utterance $y$ be represented by a vector $H^y_k(t) = [X_{1,k}, \dots, X_{t,k}, \dots, X_{M,k}]$, where $M$ denotes the sequence length. For a regression task, let the sequence targets be $L^y$, where ${L^y \in \mathbb{R}^{D}}$, where the $D$ dimensional vector $L$ denotes the output targets, for each utterance. $\overline{H}^y_k$ in eq. \ref{eq1} is obtained from $H^y_k$ by taking the mean across all the frames for utterance $y$. The cross-correlation based saliency ($CCS$) of $i^{th}$ dimension of the $k^{th}$ layer is given by: 
\vspace{-.5mm}
\newcommand\normx[1]{\left\Vert#1\right\Vert}
\begin{equation}
S_{CCS,i,k} = \abs {\frac {Cov({\overline{H}^y_{k,i}},L^y)}{\sigma_{H^y_{k,i}}\sigma_{L^y}}} + \gamma_i, \, \, \, \, where, \, \, \, \overline{H}^y_k = \frac {1}{M} \sum_{t=1}^{M} H^y_k(t)
\label{eq1}
\end{equation}

\begin{equation}
\gamma_i = \frac {1}{N-1} \sum_{j=1, j \ne i}^{N} w_j \normx {\frac {Cov({\overline{H}^y_{i,k}},{\overline{H}^y_{j,k}})}{\sigma_{{\overline{H}^y_i}}\sigma_{{\overline{H}^y_j}}}}, \, \, \, \, where, \, \, \, w_j = \normx {\frac {Cov({\overline{H}^y_j},L^y)}{\sigma_{{\overline{H}^y_j}}\sigma_{L^y}}}
\label{eq2}
\end{equation}

\noindent  
\begin{equation}
\mu_{CCS,k} = \frac {1}{D}\sum_{l=1}^D {S_{{CCS}_{k,l}}}.
    \label{eq3}
\end{equation}
$\gamma_i$ is the sum of the weighted cross-correlation between the $i^{th}$ dimension and all other dimensions, as shown in eq. \ref{eq2}.
In our experiments we have used $\mu_{CCS,k}$ given in eq. \ref{eq3} to select salient layers of a pre-trained FM, which is obtained from a randomly sampled 30K utterances in the Train1.11.

\vspace{-1mm}
\subsection{Model Training}
\vspace{-2mm}
We have trained a multi-task (dimensional and categorical) emotion recognition model. It consists of temporal convolution (kernel size of 3), followed by a 2-layered gated recurrent unit (TC-GRU) network, consisting of 256 neurons in each layer and an embedding layer of 256 neurons. The model architecture is illustrated in Fig. \ref{fig:fig1} and the model parameters are described in Appendix \ref{sec:model_parameters}. The model was trained with Train1.11 data (see Table \ref{tab:table1}), where the performance on Valid1.11 set was used for model selection and early stopping. Concordance correlation coefficient ($CCC$) \cite{lawrence1989concordance} is used as the loss function, see Appendix \ref{sec:ccc}. Models were trained with a mini-batch of 32 and a learning rate of 0.0005. 
\vspace{-2mm}
\begin{figure}[h]
\begin{minipage}[b]{1.0\linewidth}
  \centering
  \centerline{\includegraphics[width=3.1in]{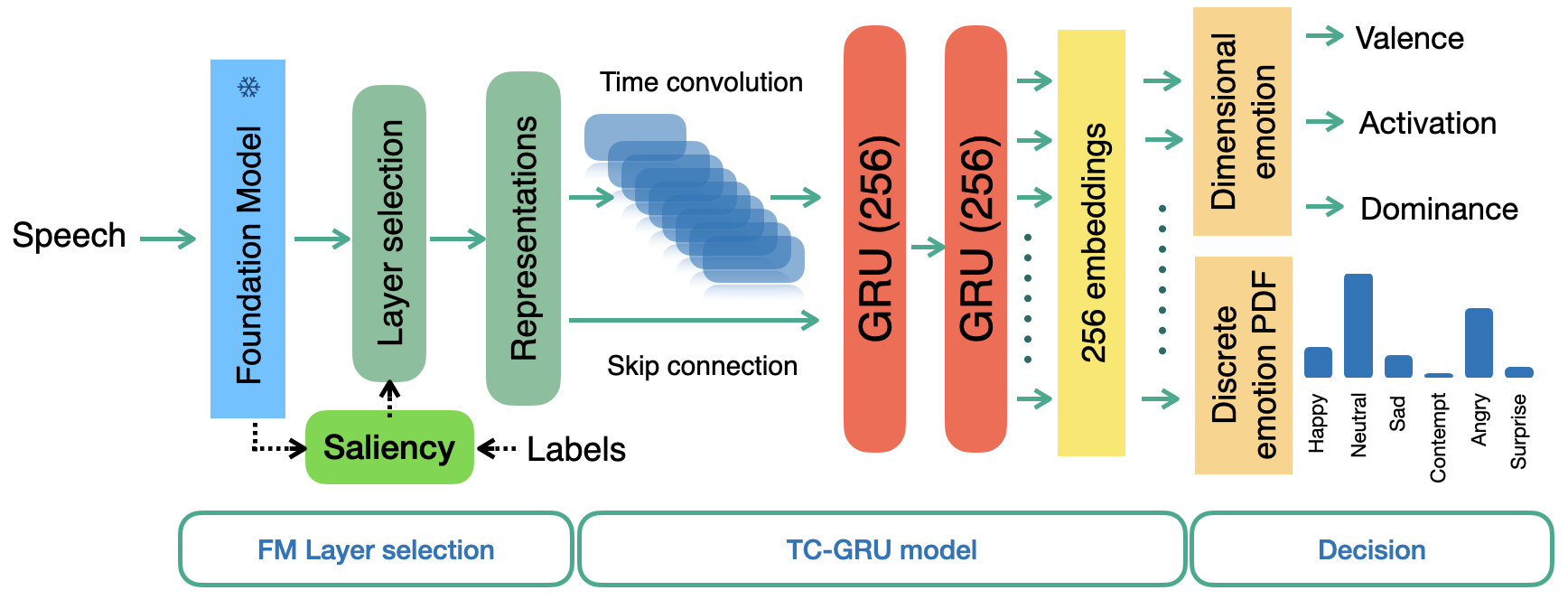}}
\end{minipage}
\vspace{-\baselineskip}
\vspace{-2mm}
\caption{Multi-task emotion recognition model}
\setlength{\belowcaptionskip}{0pt}
\label{fig:fig1}
\end{figure}
\vspace{-2mm}
\section{Results}
\vspace{-2mm}
We trained multi-task emotion recognition models with embeddings from HuBERT, WavLM, and Whisper FMs. In addition, we trained a baseline model with mel-filterbank and pitch (MFBF0) feature. In Table \ref{tab:table2}, we report dimensional emotion estimation performance obtained from the trained systems and compared them with the state-of-the-art results reported in the literature (see Table \ref{tab:table2}). Note that in \cite{srinivasan2021representation} ASR generated transcripts were used, which was not used for the other systems in Table \ref{tab:table2}. Finally, we compared categorical emotion recognition performance obtained from the TC-GRU models with respect to results reported in the literature (see Table \ref{tab:table3}). 
\vspace{-3mm}
\begin{table}[h]
\caption{Dimensional emotion estimation performance ($CCC\uparrow$) and comparison with SOTA}.
\vspace{-4mm}
\centering
\begin{adjustbox}{width=0.95\textwidth,center}
\begin{tabular}{@{}l|l @{\hspace{0.5pt}}ccc|ccc|ccc|ccc@{}}
    \toprule
    Systems & & \multicolumn{3}{c} {$\mathbf{Eval1.6}$} & \multicolumn{3}{c}{$\mathbf{Eval1.11}$} & \multicolumn{3}{c}{$\mathbf{Eval_{15dB}}$} & \multicolumn{3}{c}{$\mathbf{Eval_{5dB}}$} \\
    & &  {Act.} & {Val.} & {Dom.} & {Act.} & {Val.} & {Dom.} & {Act.} & {Val.} & {Dom.} & {Act.} & {Val.} & {Dom.} \\
    \bottomrule
    {\texttt{MFBF0 TC-GRU}} & & 0.73 & 0.34 & 0.66 & 0.62 & 0.39 & 0.56 & 0.69 & 0.26 & 0.61 & 0.53 & 0.14 & 0.48 \\
    {\texttt{HuBERT TC-GRU}} & & \textbf{0.77} & 0.65 & \textbf{0.70} & \textbf{0.66} & 0.59 & \textbf{0.59} & 0.74 & 0.62 & 0.64 & 0.61 & 0.54 & 0.49 \\ 
    {\texttt{WavLM TC-GRU}} & & \textbf{0.77} & 0.70 & \textbf{0.70} & \textbf{0.66} & 0.63 & 0.58 & 0.73 & \textbf{0.71} & \textbf{0.66} & 0.62 & 0.64 & 0.53 \\
    {\texttt{Whisper TC-GRU}} & & 0.75 & \textbf{0.71} & 0.69 & 0.65 & \textbf{0.64} & 0.58 & 0.73 & \textbf{0.71} & \textbf{0.66} & \textbf{0.66} & \textbf{0.69} & \textbf{0.60}\\
    \bottomrule
    \cite{mitra2024investigating} & & 0.75 & 0.66 & 0.67 & - & - & - & - & - & - & - & - & - \\
    \cite{srinivasan2021representation} & & \textbf{0.77} & 0.69 & 0.68 & - & - & - & - & - & - & - & - & - \\
    \bottomrule
    \end{tabular}
    \end{adjustbox}
    \label{tab:table2}
\end{table}
\vspace{-4mm}
\begin{table}[h]
\caption{Categorical emotion recognition performance and comparison with SOTA models}.
\vspace{-1mm}
\centering
\begin{adjustbox}{width=0.95\textwidth,center}
\begin{tabular}{@{}l|l @{\hspace{0.5pt}}cc|cc|cc|cc|cc|cc@{}}
    \toprule
    Systems & & \multicolumn{2}{c} {$\mathbf{Eval1.6}$} & \multicolumn{2}{c}{$\mathbf{Eval1.11}$} & \multicolumn{2}{c}{$\mathbf{Eval_{15dB}}$} & \multicolumn{2}{c}{$\mathbf{Eval_{5dB}}$} & \multicolumn{2}{c}{$Mosei$} & \multicolumn{2}{c}{$Inhouse$} \\
    & &  {$F1_m$} & {UAR} & {$F1_m$} & {UAR} & {$F1_m$} & {UAR} & {$F1_m$} & {UAR} & {$F1_m$} & {UAR} & {$F1_m$} & {UAR} \\
    \bottomrule
    {\texttt{MFBF0 TC-GRU}} & & 0.34 & 0.45 & 0.44 & 0.58 & 0.46 & 0.53 & 0.46 & 0.50 & 0.40 & 0.56 & 0.19 & 0.34 \\
    {\texttt{HuBERT TC-GRU}} & & 0.49 & 0.67 & 0.48 & 0.67 & 0.49 & 0.65 & 0.50 & 0.64 & \textbf{0.48} & 0.66 & 0.58 & 0.64 \\ 
    {\texttt{WavLM TC-GRU}} & & 0.50 & \textbf{0.70} & 0.48 & \textbf{0.69} & 0.50 & \textbf{0.69} & 0.50 & \textbf{0.68} & 0.47 & \textbf{0.69} & \textbf{0.61} & \textbf{0.67} \\
    {\texttt{Whisper TC-GRU}} & & \textbf{0.52} & 0.69 & \textbf{0.50} & 0.68 & \textbf{0.52} & 0.68 & \textbf{0.52} & 0.67 & \textbf{0.48} & 0.68 & 0.59 & 0.65 \\
    \bottomrule
    \cite{das2024speechverse} & & - & 0.67 & - & - & - & - & - & - & - & - & - & - \\
    \cite{feng2023peft} & & - & 0.67 & - & - & - & - & - & - & - & - & - & - \\
    \cite{wu2024emo} & & 0.35 & - & - & - & - & - & - & - & - & - & - & - \\
    \bottomrule
    \end{tabular}
    \end{adjustbox}
    \label{tab:table3}
\end{table}
\vspace{-3mm}

Next we investigated how these models perform across speakers, where we accumulated model decisions by speaker, and computed the UAR for the categorical emotion predictions. We have used Eval1.11 and Inhouse sets to compare the performance of the models. For performance evaluation across speakers, we introduced a metric: paUAR-X, which measures the percentage of speakers who are above a UAR of X\%, where we have used two thresholds: X: 75\% and 50\%, respectively. Table \ref{tab:table5} shows paUAR-75 and paUAR-50 for categorical emotion, obtained across speakers. Note that Tables \ref{tab:table2} and \ref{tab:table3} show that overall WavLM TC-GRU model performed better than the HuBERT TC-GRU, however table \ref{tab:table5} shows that a better system may not necessarily generalize across speakers.

\vspace{-3mm}
\begin{table}[h]
\caption{Emotion recognition performance across speakers}{where paUAR-X is the percentage of speakers who are above a UAR of X\%}.
\vspace{-1mm}
\centering
\begin{adjustbox}{width=0.93\textwidth,center}
\begin{tabular}{@{}ll|l|l @{\hspace{0.5pt}}cc|cc|cc|cc@{}}
    \toprule
    & Eval Sets & & & \multicolumn{2}{c} {$\mathbf{MFBF0\hspace{.15cm}TC-GRU}$} & \multicolumn{2}{c}{$\mathbf{HuBERT\hspace{.15cm}TC-GRU}$} & \multicolumn{2}{c}{$\mathbf{WavLM\hspace{.15cm}TC-GRU}$} & \multicolumn{2}{c}{$\mathbf{Whisper\hspace{.15cm}TC-GRU}$} \\
    & & hyps. & & {paUAR-75} & {paUAR-50} & {paUAR-75} & {paUAR-50} & {paUAR-75} & {paUAR-50} & {paUAR-75} & {paUAR-50} \\
    \bottomrule
    & {\texttt{Eval1.11}} & 1-best & & 2.3 & 21.1 & 3.1 & 34.4 & 3.1 & 40.6 & 4.7 & 39.1 \\ 
    & {\texttt{Inhouse}} & & & 2.3 & 14.0 & 10.0 & 30.5 & 9.3 & 35.0 & 11.6 & 39.5 \\ 
    \bottomrule
    & {\texttt{Eval1.11}} & 2-best & & 11.7 & 46.9 & 21.1 & 61.7 & 28.1 & 68.8 & 22.7 & 68.0 \\ 
    & {\texttt{Inhouse}} & & & 23.0 & 100.0 & 93.0 & 100.0 & 100.0 & 100.0 & 100.0 & 100.0 \\ 
    \bottomrule
    & {\texttt{Eval1.11}} & 3-best & & 20.3 & 68.0 & 36.0 & 86.7 & 46.1 & 85.9 & 36.7 & 89.8 \\ 
    & {\texttt{Inhouse}} & & & 53.1 & 100.0 & 95.4 & 100.0 & 100.0 & 100.0 & 100.0 & 100.0 \\ 
    \bottomrule
    \end{tabular}
    \end{adjustbox}
    \label{tab:table5}
\end{table}
\vspace{-1mm}

In terms of the 1-best hypothesis paUAR-75 and paUAR-50, Whisper TC-GRU model performed better than the others, likely due the fact it was pre-trained with a noisy, more diverse, and larger set of speech. However, even with this best performing model, only 5\% and 12\% of speakers had UAR above 0.75 for Eval1.11 and Inhouse sets, respectively. In Appendices \ref{sec:gender_perf} and \ref{sec:spkr_dist}, we explore potential explanations for the speaker-level performance differences including whether gender or emotion label distributions play a role. We find that gender has a significant impact on results, where $7\%$ of female speakers had UAR above 0.75 compared to $\approx 14\%$ of male speakers for the Inhouse evaluation set. This gap illustrates the importance of evaluating model performance at speaker and group levels. Interestingly, even if Tables \ref{tab:table2} and \ref{tab:table3} show that WavLM TC-GRU model overall performed much better than MFBF0 TC-GRU, their paUAR-75 were comparable for Eval1.11, indicating that the usage of overall metrics while assessing the usefulness of a model can be deceiving. Also note that the speaker level performance obtained from Eval1.11 and the inhouse set was quite different for each of the models investigated, where the performance for Eval1.11 was found to be lower, as it is a harder and larger containing more speakers than the inhouse set (see table \ref{tab:table1} in \ref{sec:data-split}). Note that for Eval1.11, the best model demonstrated an UAR above 0.75 for only $\approx 5\%$ of the speakers. The poor performance across speakers can be attributed to the uncertainty in the labels and the overall skew toward ``neutral'' emotion. For example, in many instances different graders assigned different emotions to the same speech file, which reveals that a speech file can contain a mix-of-emotions due to mixed, shifting, subtle, or ambiguous emotions. Additionally, data skew due to one emotion category being present overwhelmingly in the training set (e.g., ``neutral") can lead the model to over-estimate that emotion, in which case a 1-best hypothesis may lead to pessimistic results. Appendix \ref{sec:2hyp} illustrates the relationship between 1-best and 2-best hypothesis, and how by studying both we can obtain better clarity regarding the models generalization capacity. Table \ref{tab:table5}, we explored paUAR-X if the target emotion exists within the 2-best or 3-best hypotheses. We find a paUAR-75 of more than 28\% can be obtained by considering the 2-best hypothesis and as high as 46\% can be obtained with a 3-best hypothesis. These findings indicate that (1) in case of data with uncertain labels and distribution skew, it is helpful to consider multiple model hypothesis and (2) label distribution skew impacts model's generalization capacity across speakers.

\vspace{-2mm}
\section{Conclusions}
\vspace{-3mm}
In this work, we demonstrated SOTA results for both dimensional and categorical emotion recognition. The models were found to perform well for unseen datasets (Mosei and Inhouse) and demonstrated reasonable noise robustness. Interestingly, the models failed to generalize across speakers, where we observed that the model performed with an overall UAR of above 0.75 for less than $10\%$ of the speakers. The model offered UAR above 0.5 for $\approx60\%$ of the speakers. This indicated that using metrics that reflect the overall performance on an eval set may not be prudent, speaker-level and gender-level performance are crucial to assess how well the model will perform across users. We also observed that instead of using the 1-best hypothesis from the model, it is useful to consider 2-best or 3-best hypothesis, as certain utterances may contain multiple emotions, in which case the model may provide more than one likely emotion categories. With 2-best and 3-best hypothesis, we observed that UAR above 0.75 was obtained for $>60\%$ and $>85\%$ of the speakers, respectively. The findings from this study opens the question regarding performance metrics, which can account for co-occurrences of semantically closer emotions, such as ``angry", ``contempt", ``disgust", which may have a higher chance of confusion with each other.

\bibliography{iclr2025}
\bibliographystyle{iclr2025}

\appendix
\section{Appendix}

\subsection{Concordance Correlation Coefficient}
\label{sec:ccc}

Concordance correlation coefficient based loss ($L_{ccc}$) is defined by:
\begin{equation}
L_{ccc} = - \frac{1}{N} \sum_{i=1}^{N} CCC_{i}
  \label{eq4}
\end{equation}
where ${L_{ccc}}$ is the mean of $CCC$'s obtained from each of the $N$ output targets. $CCC$ is defined by:
\begin{equation}
CCC = \frac {2\rho \sigma_x\sigma_y}{\sigma_x^2+\sigma_y^2 +(\mu_x-\mu_y)^2 }.
    \label{eq5}
\end{equation}
where ${\mu _{x}}$ and ${\mu _{y}}$ are the means, ${\sigma _{x}^{2}}$ and ${\sigma _{y}^{2}}$ are the corresponding variances for the estimated and groundtruth variables, and ${\rho}$ is the correlation coefficient between them.

\subsection{Data split}
\label{sec:data-split}
\begin{table}[ht]
\centering
\caption{MSP-podcast data split, noise-degraded test sets and out-of-domain MOSEI and Inhouse evaluation set}
\vspace{2mm}
\begin{tabular}{llll}
\hline\hline
\textbf{Split} & \textbf{Hours} & \textbf{Speakers} & \textbf{Description}\\
\hline\hline
MSP Train1.11 & 135.4 & 1411 & Training set \\
MSP Valid1.11 &  31.7 & 456 & Validation set \\
\hline
MSP Eval1.6 &  16.6 & 51 & Podcast1.6 evaluation set \\
MSP Eval1.11 &  48.9 & 244 & Podcast1.11 Eval set 1 \\
MSP $Eval_{15dB}$ & 28.4 & 51 & Eval + noise within 10-20 dB \\
MSP $Eval_{5dB}$ & 28.4 & 51 &Eval + noise within 0-10 dB \\
CMU-Mosei &  70.6 & - & Mosei segments  \\
Inhouse data & 5.0 & 85 & Conversational speech segments  \\
\hline\hline
\end{tabular}
\label{tab:table1}
\end{table}

\subsection{Layer Saliency measure}
\label{sec:Saliency}

Neural saliency was used in \cite{mitra2024investigating} to reduce the number of representations for the downstream task with a goal of model size reduction. “Saliency” in this work focuses on layer-saliency as outlined in section \ref{sec:rep}, where the saliency measure was modified to provide a layer-wise collective measure, that informs which transformer layer in the foundation model is more relevant. This measure is particularly important, as given the large number of transformer layers in an FM, it may not be possible to perform layer-wise experimentation of which layer offers the best representation. Layer-wise saliency measure offers a data-driven solution to figure out which layers in the transformer network are better suited for the downstream task, without the need to train downstream models for representations from each individual layer. 

\begin{figure}[H]
\begin{minipage}[b]{1.0\linewidth}
  \centering
  \centerline{\includegraphics[width=4in]{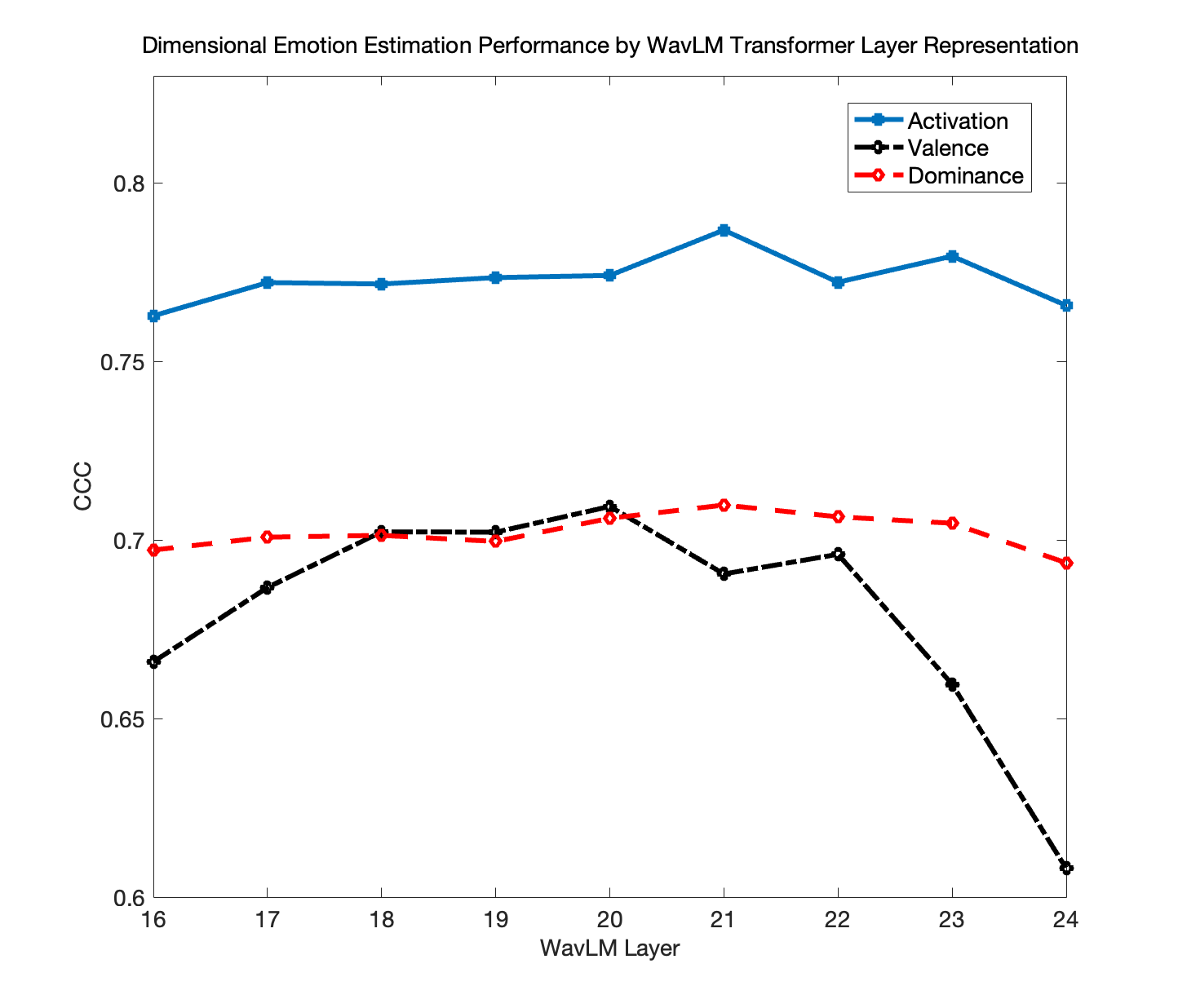}}
\end{minipage}
\vspace{-\baselineskip}
\caption{Dimensional emotion estimation for different transformer layers in WavLM}
\setlength{\belowcaptionskip}{0pt}
\label{fig:fig6}
\end{figure}

\begin{figure}[H]
\begin{minipage}[b]{1.0\linewidth}
  \centering
  \centerline{\includegraphics[width=4in]{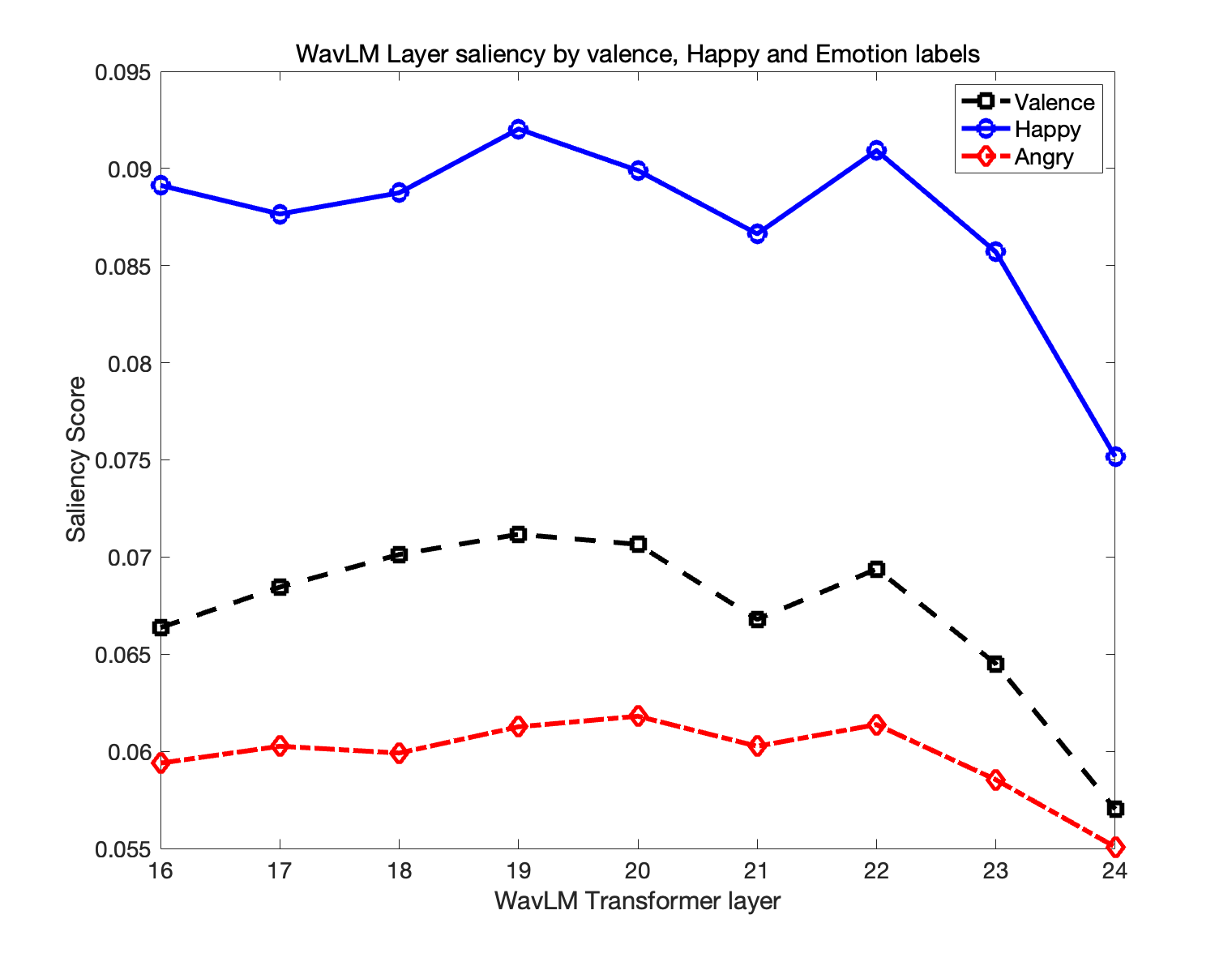}}
\end{minipage}
\vspace{-\baselineskip}
\caption{WavLM layer saliency by valence, happy and angry emotion}
\setlength{\belowcaptionskip}{0pt}
\label{fig:fig7}
\end{figure}

We observed that valence is more sensitive to transformer layer representation, compared to activation and dominance (see Figure \ref{fig:fig6}). Earlier studies \cite{chen2022wavlm} have found that for WavLM intermediate layers (specifically layers 19 and 20) are better for intent classification. Valence plays an important role in emotion discrimination, such as Happy versus Angry or Sad versus Calm. In Figure \ref{fig:fig7} we show how saliency based on individual valence, happy and angry scores vary by WavLM transformer layer representation. Figures \ref{fig:fig6} and \ref{fig:fig7} show that intermediate transformer layers of WavLM offer better representations (paralinguistic cues) for downstream emotion detection compared to the final layer. We observed that the intermediate layers correlated strongly with articulatory features (extracted using the model in \cite{mitra2018articulatory}), speech rate, pitch and voicing information, compared to the final layer. 

\subsection{Emotion Model details}
\label{sec:emotion_modeling}

Table \ref{tab:table6} shows that representations from emotion-salient layer as compared to the final FM layer, resulted in improvement in emotion recognition performance. It is also interesting to note that relative improvement in valence was higher ($> 8\%$ relative) compared to the other dimensional emotions. For unseen-noise sets ($Eval_{15dB}$ and $Eval_{5dB}$), the relative improvement was higher (16.5\% for dimensional and 10\% for categorical emotion) than other evaluation sets. \footnote{performance gains from the salient FM-layer representations were statistically significant ($p < 0.05$) compared to the results reported in the literature}

\begin{table}[H]
\caption{Dimensional and categorical emotion estimation using (1) MFBF0 feature, (2) FM representations from final layer and (3) FM representations from the salient layer}.
\vspace{2mm}
\centering
    \begin{adjustbox}{width=0.7\textwidth,center}
        \begin{tabular}{@{}lll@{\hspace{0.5pt}}ccccc@{}}
            \toprule
             Test set & Reps. & Layer & \multicolumn{3}{c}{Dim. Emo.} & \multicolumn{1}{c}{Cat. Emo.}& \\
             & & & & {CCC $\uparrow$} & & {UAR $\uparrow$} \\
             & & & {Act.} & {Val.} & {Dom.} & \\

            \toprule
            \multirow{6}{*}{$\mathbf{Eval1.3}$} & \multicolumn{1}{l}{\texttt{MFBF0}} & \multicolumn{1}{l}{$-$} & 0.73 & 0.33 & 0.67 & 0.55 \\\cline{2-7}
            
            & \multicolumn{1}{l}{\texttt{HuBERT}} & \multicolumn{1}{l}{Final} & 0.75 & 0.60 & 0.69 & 0.65\\
            
            & & \multicolumn{1}{l}{Salient} & \textbf{0.78} & 0.65 & \textbf{0.71} & 0.66 \\\cline{2-7}
            
            & \multicolumn{1}{l}{\texttt{WavLM}} & \multicolumn{1}{l}{Final} & 0.77 & 0.61 & 0.70 & 0.66 \\
            
            & & \multicolumn{1}{l}{Salient} & 0.77 & 0.70 & \textbf{0.71} & \textbf{0.71} \\\cline{2-7}

            & \multicolumn{1}{l}{\texttt{Whisper}} & \multicolumn{1}{l}{-} & 0.76 & \textbf{0.71} & 0.69 & \textbf{0.71} \\
                        
            \bottomrule

            \multirow{6}{*}{$\mathbf{Eval1.6}$} & \multicolumn{1}{l}{\texttt{MFBF0}} & \multicolumn{1}{l}{$-$} & 0.73 & 0.34 & 0.66 & 0.54 \\\cline{2-7}
            
            & \multicolumn{1}{l}{\texttt{HuBERT}} & \multicolumn{1}{l}{Final} & 0.75 & 0.60 & 0.68 & 0.64\\
            
            &  & \multicolumn{1}{l}{Salient} & \textbf{0.77} & 0.65 & \textbf{0.70} & 0.66 \\\cline{2-7}
            
            & \multicolumn{1}{l}{\texttt{WavLM}} & \multicolumn{1}{l}{Final} & 0.76 & 0.61 & 0.69 & 0.65 \\
            
            &  & \multicolumn{1}{l}{Salient} & \textbf{0.77} & 0.70 & \textbf{0.70} & \textbf{0.70} \\\cline{2-7}

            & \multicolumn{1}{l}{\texttt{Whisper}} & \multicolumn{1}{l}{-} & 0.75 & \textbf{0.71} & 0.69 & 0.61 \\
            
            \bottomrule
            
            \multirow{6}{*}{$\mathbf{Eval1.11}$} & \multicolumn{1}{l}{\texttt{MFBF0}} & \multicolumn{1}{l}{$-$} & 0.62 & 0.39 & 0.56 & 0.58 \\\cline{2-7}
            
            & \multicolumn{1}{l}{\texttt{HuBERT}} & \multicolumn{1}{l}{Final} & 0.64 & 0.55 & 0.57 & 0.65\\
            
            &  & \multicolumn{1}{l}{Salient} & \textbf{0.66} & 0.59 & \textbf{0.59} & 0.67\\\cline{2-7}
            
            & \multicolumn{1}{l}{\texttt{WavLM}} & \multicolumn{1}{l}{Final} & 0.65 & 0.57 & 0.58 & 0.65 \\
            
            & & \multicolumn{1}{l}{Salient} & \textbf{0.66} & 0.63 & 0.58 & \textbf{0.69 }\\\cline{2-7}

            & \multicolumn{1}{l}{\texttt{Whisper}} & \multicolumn{1}{l}{-} & 0.65 & \textbf{0.64} & 0.58 & 0.68 \\
            \bottomrule
            \bottomrule

            \multirow{6}{*}{$\mathbf{Eval_{15dB}}$} & \multicolumn{1}{l}{\texttt{MFBF0}} & \multicolumn{1}{l}{$-$} & 0.69 & 0.26 & 0.61 & 0.50 \\\cline{2-7}
            
            & \multicolumn{1}{l}{\texttt{HuBERT}} & \multicolumn{1}{l}{Final} & 0.70 & 0.57 & 0.60 & 0.62\\
            
            & & \multicolumn{1}{l}{Salient} & \textbf{0.74} & 0.62 & 0.64 & 0.64\\\cline{2-7}
            
            & \multicolumn{1}{l}{\texttt{WavLM}} & \multicolumn{1}{l}{Final} & 0.72 & 0.58 & 0.62 & 0.63 \\
            
            & & \multicolumn{1}{l}{Salient} & 0.73 & \textbf{0.71} & \textbf{0.66} & \textbf{0.68} \\\cline{2-7}

            & \multicolumn{1}{l}{\texttt{Whisper}} & \multicolumn{1}{l}{-} & 0.73 & \textbf{0.71} & \textbf{0.66} & 0.67 \\
            \bottomrule

            \multirow{6}{*}{$\mathbf{Eval_{5dB}}$} & \multicolumn{1}{l}{\texttt{MFBF0}} & \multicolumn{1}{l}{$-$} & 0.53 & 0.14 & 0.48 & 0.44 \\\cline{2-7}
            
            & \multicolumn{1}{l}{\texttt{HuBERT}} & \multicolumn{1}{l}{Final} & 0.56 & 0.49 & 0.43 & 0.56\\
            
            & & \multicolumn{1}{l}{Salient} & 0.61 & 0.54 & 0.49 & 0.60\\\cline{2-7}
            
            & \multicolumn{1}{l}{\texttt{WavLM}} & \multicolumn{1}{l}{Final} & 0.61 & 0.50 & 0.48 & 0.60 \\
            
            & & \multicolumn{1}{l}{Salient} & 0.62 & 0.64 & 0.53 & \textbf{0.66} \\\cline{2-7}

            & \multicolumn{1}{l}{\texttt{Whisper}} & \multicolumn{1}{l}{-} & \textbf{0.66} & \textbf{0.69} & \textbf{0.60} & 0.65 \\
            
            \bottomrule
            \multirow{6}{*}{$\mathbf{Mosei}$} & \multicolumn{1}{l}{\texttt{MFBF0}} & \multicolumn{1}{l}{$-$} & - & - & - & 0.56 \\\cline{2-7}
            
            & \multicolumn{1}{l}{\texttt{HuBERT}} & \multicolumn{1}{l}{Final} & - & - & - & 0.64\\
            
            & & \multicolumn{1}{l}{Salient} & - & - & - & 0.66\\\cline{2-7}
            
            & \multicolumn{1}{l}{\texttt{WavLM}} & \multicolumn{1}{l}{Final} & - & - & - & 0.66 \\
            
            & & \multicolumn{1}{l}{Salient} & - & - & - & \textbf{0.69} \\\cline{2-7}

            & \multicolumn{1}{l}{\texttt{Whisper}} & \multicolumn{1}{l}{-} & - & - & - & 0.68 \\
            \bottomrule

        \end{tabular}
    \end{adjustbox}
    \label{tab:table6}
\end{table}

\subsection{Performance by Gender}
\label{sec:gender_perf}

Table \ref{tab:table8} shows performance variance across male and female speakers for Eval1.11 and Inhouse test sets. We find performance is considerably lower for female speakers across both datasets, and the gap between performance for male and female speakers increases with the paUAR threshold. The training set is skewed toward male speakers, which likely contributes to the observation in Table \ref{tab:table8}.

\begin{table}[h]
\caption{Emotion recognition performance (paUAR-75 and paUAR-50) by gender for Whisper TC-GRU model}.
\centering
\begin{tabular}{@{}ll| cc|cc@{}}
    \toprule
    Eval Sets & & \multicolumn{2}{c} {paUAR-75} & \multicolumn{2}{c}{paUAR-50} \\
    & &  {Female} & {Male} & {Female} & {Male} \\
    \bottomrule
    {\texttt{Eval1.11}} & & 2.7 & 7.1 & 45.8 & 50.4 \\
    {\texttt{Inhouse}} & & 7.1 & 13.8 & 28.6  & 44.8 \\
    \bottomrule

    \end{tabular}
    \label{tab:table8}
\end{table}

\subsection{Performance by Speaker's Emotion Distributions}\label{sec:spkr_dist}

\begin{figure}[H]
    \centering
    \includegraphics[width=1.\linewidth]{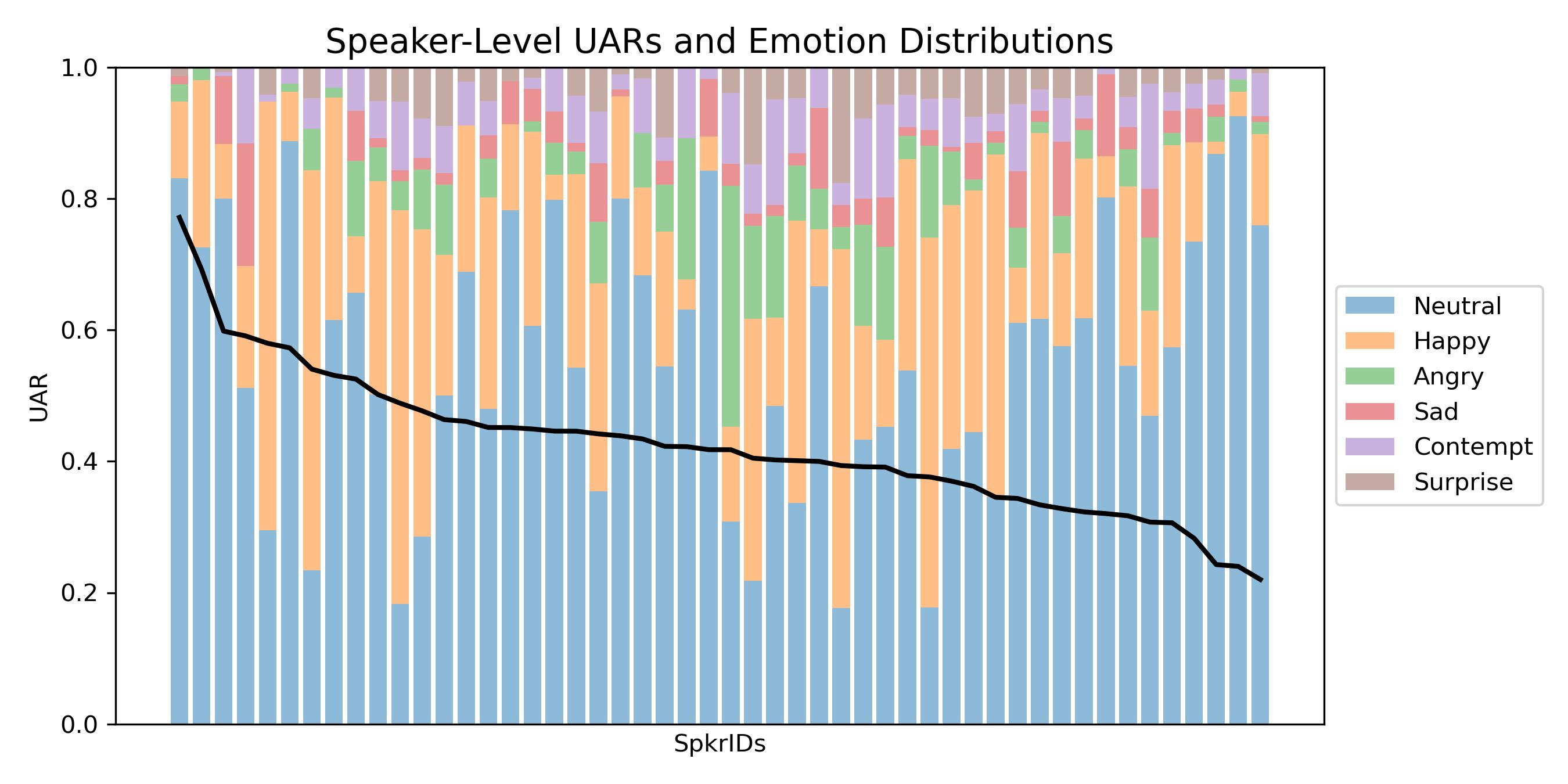}
    \vspace{-3mm}
    \caption{Speaker-level performance (UAR from Whisper TC-GRU) plotted against emotion distributions,  for speakers in Eval1.6.}
    \label{fig:spkr_dists}
\end{figure}

Figure \ref{fig:spkr_dists} shows performance plotted against emotion distributions for each speaker in Eval1.6. Because UAR is the unweighted average across recall on all emotions, we do not find a strong relationship between UAR and emotion distribution. This suggests UAR is robust to these speaker-level changes and can capture other important factors in speaker-level performance. 

\subsection{Relationship Between 1st and 2nd Best Model Hypotheses}\label{sec:2hyp}
\begin{figure}
    \centering
    \includegraphics[width=\linewidth]{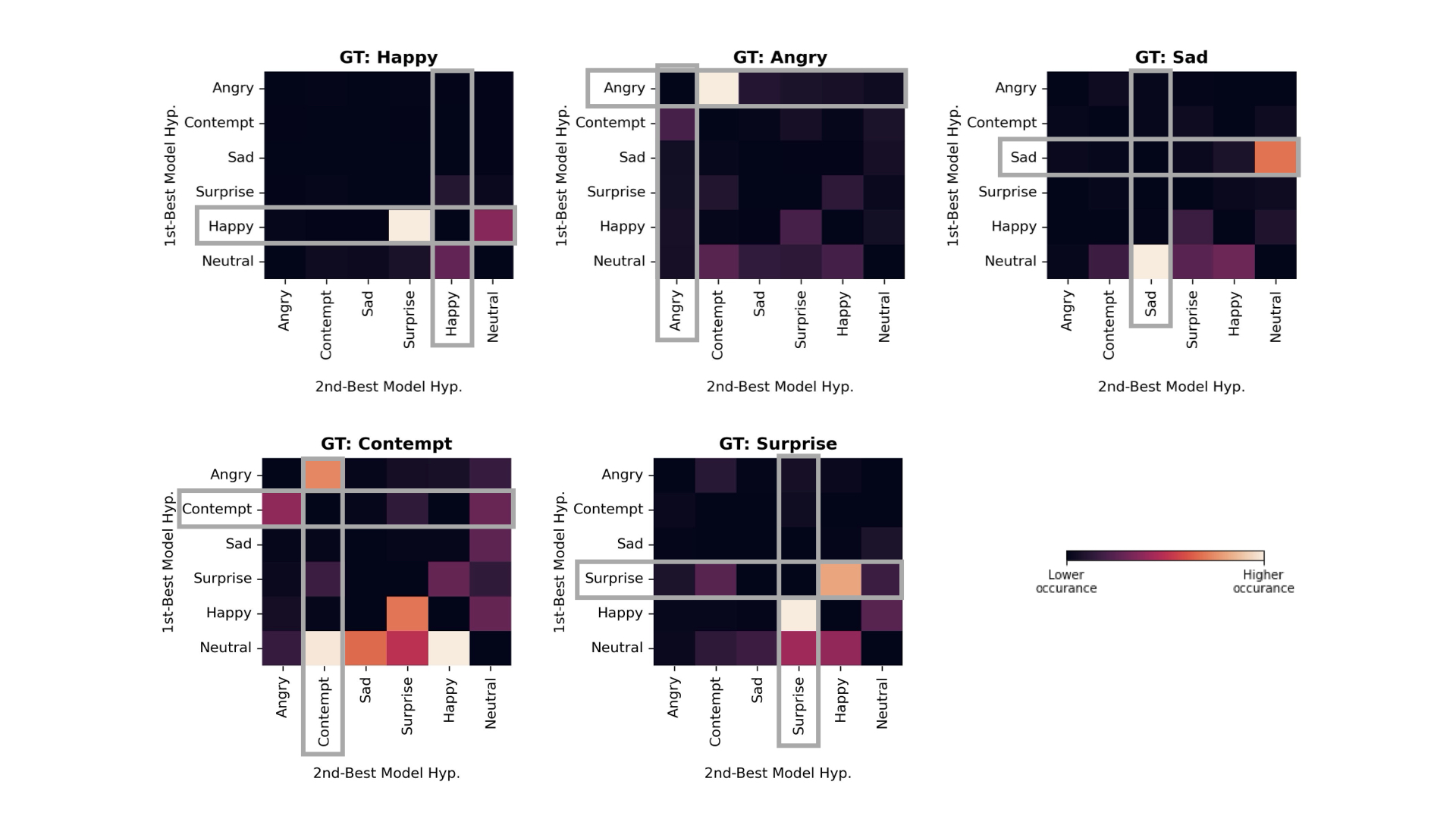}
    \vspace{-10mm}
    \caption{Confusion matrices showing the relationship between 1st and 2nd best model hypotheses from Whisper TC-GRU and the Eval1.6 test set.}
    \label{fig:hyp_cms}
\end{figure}

We find that the model's first and second hypotheses show a clear relationship, and that the first hypothesis alone may not fully reflect the model's understanding. Figure \ref{fig:hyp_cms} illustrates these details, with the samples accurately labeled by the first hypothesis outlined by the horizontal gray bars, and the samples accurately labeled by the second hypothesis outlined by the vertical gray bars. The first hypotheses are highly accurate for happiness and anger, indicated by the white squares within the horizontal gray outlines. However, for most sadness samples, the model identifies neutral as the most likely emotion and sadness as the second most likely emotion, indicated by the white square within the vertical gray outline. Similarly, for surprise samples, the model identifies happiness as the most likely emotion and surprise as the second most likely emotion, where this hierarchy likely results from the closer relationship between happiness and surprise with the former class having more representation in the training data. We also see considerable confusion between contempt, anger, and neutral. When we explore the models second best hypotheses, we find the model correctly detects the overall sentiment but does not distinguish correctly between them. This finding supports our analysis into considering the model's second best hypotheses when determining model predictions.

\subsection{Model Parameters}
\label{sec:model_parameters}

The TC-GRU models had 1.6M parameters (2.1MB), whereas the MFBF0 was 700KB in size, for saliency based layer selection, we were able to reduce the computation needed by WavLM (16\%) and by HuBERT (8\%) by reducing the number of transformer layers needed to generate the representations, see Table \ref{tab:table7}. Note that layers were all frozen for feature extraction, i.e., none of the FM transformer layers were fine-tuned for the given task as shown in Figure \ref{fig:fig1}. Earlier work \cite{mitra2024investigating} has shown that saliency-based representation selection can help to reduce the downstream model size, however that was not the focus of this work. The goal of this work is to investigate layers that are relevant for downstream emotion task, where joint modeling of categorical and dimensional emotion would result in better performance, as compared to using the final layers. Note that most studies have used FM final layer representations to train teacher models to distill information into simpler downstream models, in this work we show that better teacher models can be obtained by proper selection of representation layers.  
\newcolumntype{L}{>{\centering\arraybackslash}m{3cm}}
\newcolumntype{M}{>{\centering\arraybackslash}m{2.5cm}}
\newcolumntype{N}{>{\centering\arraybackslash}m{2cm}}
\addtolength{\tabcolsep}{-0.6em} 
\begin{table}[ht]
\centering
\caption{Model Parameters}
\vspace{2mm}
\begin{tabular}{LMMNNN}
\hline\hline
\textbf{Model} & \textbf{Full FM Params} (Not loaded) & \textbf{Salient Layer FM Params} (Loaded, frozen) & \textbf{Saliency Size Reduction} & \textbf{Trainable Params} (TC-GRU) & \textbf{Total Loaded Model Size} \\
\hline\hline
MFBF0 TC-GRU & - & - & - & 0.5M & 0.7MB \\
WavLM TC-GRU & 315.5M & 265.1M & 16\% & 1.6M & 2.2MB \\
HuBERT TC-GRU &  315.4M & 290.2M & 8\% & 1.6M & 2.2MB\\
Whisper TC-GRU & 315.7M & 315.7M  & 0\% & 1.6M & 2.2MB \\
\hline\hline
\end{tabular}
\label{tab:table7}
\end{table}

\end{document}

%% file: math_commands.tex
\usepackage{amsmath,amsfonts,bm}

\def\eqref#1{equation~\ref{#1}}

\def\1{\bm{1}}

\DeclareMathAlphabet{\mathsfit}{\encodingdefault}{\sfdefault}{m}{sl}
\SetMathAlphabet{\mathsfit}{bold}{\encodingdefault}{\sfdefault}{bx}{n}

